\documentclass[11pt]{article}
\usepackage{titlesec}
\usepackage{amsmath,esdiff}
\usepackage{amssymb}
\usepackage{authblk}

\usepackage[normalem]{ulem}
\usepackage{bbold}
\usepackage{BOONDOX-cal}
\usepackage[colorlinks=true,linkcolor=blue,citecolor=teal,urlcolor=teal]{hyperref}%
\usepackage{geometry}

\usepackage{mathtools}
\usepackage{subcaption} 
\usepackage{url}
\usepackage{doi}
\usepackage{xcolor}

\definecolor{mmaBlue}{RGB}{54, 92, 141}
\definecolor{mmaOrange}{RGB}{229, 119, 40}
\definecolor{mmaGreen}{RGB}{126, 167, 60}

\usepackage{diagbox} 

\usepackage{setspace}
\usepackage{cite}
\usepackage{flushend}
\usepackage{mathrsfs}
\usepackage{hyperref}
\usepackage{graphicx}
\usepackage{cancel}
\usepackage{balance}
\usepackage{array,esint}
\usepackage[most]{tcolorbox}
\graphicspath{}
\titleformat{\section}{\fontsize{12}{12}\bfseries}{\thesection}{1em}{}
\twocolumn
\begin{document}
\twocolumn[\begin{@twocolumnfalse}
\title{\textbf{Revisiting Thermodynamics of the Hayward Black Holes and Exploring Binary Merger Bounds}}
\author{\textbf{Neeraj Kumar${}^{a ~*}$, Ankur Srivastav${}^{b~\ddagger}$, Phongpichit Channuie${}^{a~ \dagger}$}}
\affil{{${}^{a}$ School of Science, Walailak University}\\
{Nakhon Si Thammarat, 80160, Thailand}\\
{${}^{b}$ Vahrenwalder Str., Hannover-30165, Germany}\\
{}}
\date{}
\maketitle

\begin{abstract}
\noindent In this article, we revisit the thermodynamics of Hayward black holes \cite{Hayward} in asymptotic flat spacetime and obtain the bounds on the final mass post merger after head-on collision event of two equal mass black holes. We revisit thermal properties of these black holes from a perspective that the laws of black hole thermodynamics remain valid. Under this condition a novel entropy formula appears naturally with a logarithmic correction term along with one extra term. We discuss phase structure of the black holes. Next, we obtain the bounds on the final black hole mass parameter using the validity of the second law of black hole thermodynamics with the new entropy formula. We discuss the impact of the Hayward parameter on thermal and merger properties of these black holes. 
\end{abstract}
\end{@twocolumnfalse}]

\noindent\let\thefootnote\relax\footnote{{}\\
{$*$nkneeraj06@gmail.com}\\
{$\ddagger$ankursrivastavphd@gmail.com}\\
{$\dagger$}phongpichit.ch@mail.wu.ac.th}

\section{Introduction}

Black holes are the most fascinating solutions in gravity which predict the breakdown of a classical theory, thus, providing a window for a discovery of a complete theory which could explain the associated spacetime singularity. A few candidate theories of quantum gravity have been proposed to deal with the issue of classical singular solutions. However, many ambiguities are still present in these approaches \cite{kiefer01, kiefer02}. On the other hand, there have been progress in phenomenologically finding a cure to these singular solutions by introducing new parameters such that these provide a proxy for the high energy effects. These phenomenological approaches are motivated in the spirit of effective field theory prescription in particle physics. Bardeen \cite{Bardeen} proposed the first regular black hole solution which was also shown later to be a solution of Einstein's gravity coupled to non-linear electrodynamics \cite{ayon1}. In 2005, Hayward \cite{Hayward} proposed a model for regular black holes, motivated from the physical perspective of the black hole formation and evaporation processes. Unlike Bardeen black hole, even though it cannot be derived from a fundamental Lagrangian, this model is much simple. Also, the casual structure remains similar to the Bardeen black hole. The Hayward black hole has Schwarzschild-like geometry at spacelike infinity and de-Sitter-like geometry near the core. The black hole has two horizons for a range of parameters and these two merge at extremal limit. Interesting cases of the geodesic motion outside the event horizon \cite{geodesic} and  the quasi-normal mode analysis \cite{qnm} for these black holes have been explored in the literature before.\\ 
\noindent Existence of event horizon makes the black hole geometry interesting from thermodynamic point of view as well. This perspective of the Hayward black holes has also been explored in the literature \cite{martin1, martin2, Naseer}. In these analysis, a temperature expression, different from the Hawking temperature, is considered in order to comply with the first law and the Bekenstein entropy formula. Also, it is well known that there should be modifications to the entropy formula due to quantum corrections. These modifications appear as logarithmic corrections to the standard area term \cite{dabholkar, solodukhin}. Thus, one would like to consider the consistency of the first law without choosing any entropy formula apriori. Under this renewed thermal phase space, it would be interesting to study the thermal stability and phase transition structure for these black holes.  \\
\noindent One important property of the second law of thermodynamics is that it provides a constraint on binary black hole merger in light of gravitational wave analysis \cite{Hawkingmerger}. With the advent of the multi-messenger astronomy, the data from such merger events are becoming more available. Thus, with the data advantage in hand, it shall be interesting to test theories beyond general relativity. One of the insightful test would be to constraint phenomenological parameters for modified gravity theories.   \\
\noindent We start with a brief introduction to the Hayward black hole and discuss its causal structure along with the locations of all its horizons in section 2. Next, we consider the renewed thermodynamic phase space for these black holes and calculate the new entropy formula consistent with the first law. In this section, we also discuss the subtleties around phase transitions in these black holes. Section 4 is devoted to utilising the new entropy formula and the second law of thermodynamics to discuss a binary black hole merger scenario. The impact of the Hayward parameter on the post-merger mass parameter is studied in this section. The results and concluding remarks are given in the last section 5.

%\RepAS{}{(3) Regular BH Same thing. (4) Check \mbox{$S_n$} inequalities for \mbox{$\Lambda=0$}. (1 Done) Extension to AdS Regularization (2 Done) Black hole merger with regular entropy and variation with parameter a.} 

\section{Hayward Black Holes}

\noindent  We start with the Hayward regularized metric for an uncharged black hole in flat spacetime. This metric is given by \cite{Hayward}, 
\begin{equation}
    ds^2=-f(r)dt^2+\frac{dr^2}{f(r)}+r^2d\Omega^2,
    \label{geometry}
\end{equation}
where
\begin{equation}
    f(r)=1-\frac{2M r^2}{r^3+2M l^2}~.
    \label{eq2}
\end{equation}
Here, \(M\) is the mass parameter and \(d\Omega^2\) is a metric on a unit 2-sphere ($S^2$). Here, \(l\) is the Hayward regularization parameter which has a dimension of length. The lapse function recovers the form of an uncharged Schwarzschild black hole in \(l\rightarrow 0\) limit. In this limit, parameter \(M\) reduces to the standard ADM mass of the black hole. \\
\noindent The black hole horizon is given by \(f(r_h)=0\), which takes the following form,
\begin{equation}
    r_h^3-2Mr_h^2+2M l^2=0~.
    \label{eqn}
\end{equation}
\noindent Notice that the horizon structure is similar to the case of Reissner-Nordström black hole. The above cubic equation either has two, one or no positive real roots depending on the value of \(M\). A single root corresponds to an extremal black hole. The parameters quantifying it can be calculated at the extremum of eq.(\ref{eqn}). The horizon radius \(r_{ext}\) and the mass parameter \(M_{ext}\) for the extremal black hole are then given as,
\begin{equation}
    M_{ext}=\frac{3\sqrt{3}}{4}l,~~~~~~~~~r_{ext}=\sqrt{3}l~.
\end{equation}
For values below \(M_{ext}\) there is no horizon and hence no black hole solution. For values above \(M_{ext}\), there exist two horizons; an event horizon \(r_+\) and a Cauchy horizon \(r_-\), which is inside the event horizon. These two horizons can be expressed in terms of the parameters \(M\) and \(l\) in the following form\cite{martin1},
\begin{equation}
\begin{aligned}
    r_+ &=\dfrac{2M}{3}\left[1+2\cos{\left(\frac{1}{3}\cos^{-1}{\left(1-\frac{27l^2}{8M^2}\right)}\right)}\right]~,\\
      r_- &=\dfrac{2M}{3}\left[1+2\cos{\left(\frac{1}{3}\cos^{-1}{\left(1-\frac{27l^2}{8M^2}\right)}-\frac{2\pi}{3}\right)}\right]
      \end{aligned}
\end{equation}

\noindent Now that we have the information about the causal structure of the Hayward black hole spacetime, we next move to study its thermal properties.

\section{Thermodynamic Properties and Phase Structure}

\noindent In this work, we shall consider the case with two real roots of eq.(\ref{eqn}), i.e. when there exist two horizons for the Hayward black hole. In this section, we shall explore the thermal properties associated with the event horizon, $r_+$. We \textit{assume} that asymptotic observers observe Hawking radiation associated with the event horizon. Notice that it is one of the key assumptions of this work and it leads to a non-trivial modification to the entropy formula associated with the Hayward black hole. Such an assumption has already been considered in literature \cite{Alice01, NK3}.\ 

\noindent The Hawking temperature here can be calculated from the lapse function, given in eq.(\ref{eq2}), as 
\begin{align}
    T&=\frac{f'(r)|_{r=r_+}}{4\pi}\nonumber\\
    &=\frac{1}{4\pi}\left[\frac{1}{r_+}-\frac{3l^2}{r_+^3}\right]~.
    \label{temp}
\end{align}

\noindent Also, using eq.(\ref{eqn}), one can express the parameter \(M\) in terms of the horizon radius, \(r_+\), as
\begin{equation}
    M=\frac{r_+^3}{2(r_+^2-l^2)}~.
    \label{mass}
\end{equation}

\noindent The above expressions reduce to the Hawking temperature and the ADM mass of a Schwarzschild black hole in the limit \(l\rightarrow0\). \\
\noindent Now that we have these thermal variables, one can \textit{assume} that the first law of black hole still holds and the parameter \(M\) plays a role of the internal energy of the black hole. These assumptions are not obvious as the Hayward black holes are not solutions to the Einstein's field equations. These black holes are motivated by the quantum corrections, which are phenomenologically incorporated into the metric. However, the validity of the thermodynamics can still be assumed in the present scenario and the first law may be given as, 
\begin{equation}
    dM=TdS~.
    \label{first_law}
\end{equation}
Under these assumptions, the entropy associated with this regular black hole spacetime does not follow Bekenstein-Hawking formula (\(S=A/4\)). This is because such an entropy formula would be incompatible with eq.(\ref{first_law}). Thus, one can use this relation to find the entropy of the Hayward black hole spacetime using following integral,
\begin{equation}
    S=\int_0^{r_+}\frac{dM(r_+')}{dr_+'}\frac{1}{T}dr_+'~. 
\end{equation}
The expression for the Hayward black hole then takes the following form,
\begin{equation}
    S=\pi\left(r_+^2+2l^2\log{\frac{r_+^2-l^2}{l^2}}-\frac{l^2(r_+^2-l^2 +1)}{r_+^2-l^2}\right)~.
    \label{new_ent}
\end{equation}
This expression admits two correction terms to the standard area term. The logarithmic correction term has already been discussed in the literature \cite{dabholkar, solodukhin} as originated due to quantum corrections. These corrections are usually prominent near extremal limit, however, for the Hayward black hole, these appear automatically even in sub-extremal case. Apart from this logarithmic correction term, there is an extra term, which is inversely proportional to the horizon area. This newly discovered entropy formula, eq.(\ref{new_ent}), would be very illuminating to the black hole characteristics.  We shall elaborate upon this point in section 4.  \

\begin{figure}
    \includegraphics[width=1\linewidth]{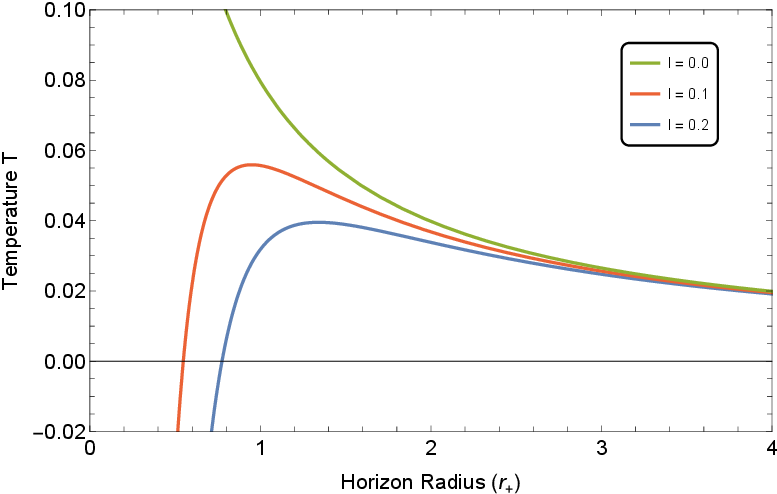}
        \caption{Hawking temperature vs horizon radius for Hayward black holes for \(l=(0, 0.1, 0.2)\). }
\label{f1}
    \end{figure}

\noindent In Fig.{(\ref{f1})}, we have plotted the black hole temperature, $T$, corresponding to the horizon radius, $r_+$, for different values of the Hayward parameter, $l$.   
The monotonically decreasing curve in the entire domain of the horizon radius corresponds to the Schwarzschild black hole (that is, \(l=0\)). However, for non-zero values of the Hayward parameter, \(l\), the behaviour changes significantly. Now, there is a non-zero horizon radius for which the temperature is zero and, hence, there is an extremal black hole limit. Another important thing to note here is that for non-zero values of the Hayward parameter, \(l\), the temperature initially rises and then falls asymptotically. It shows that the small black holes are thermally stable in comparison to large black holes in canonical ensemble. This behaviour is in contrast to the Schwarzschild black holes in AdS spacetime, where only the large black holes are stable\cite{HawkingPage}. Before moving to the stability analysis for these black holes, we want to discuss the role of dimension-full variable, \(l\) in the first law.  \\
\noindent In recent studies \cite{kostor, SG, mann, dolan}, it has been pointed out that in order to have a consistent Euler relation, which is also known as the Smarr relation \cite{smarr}, between thermodynamic variables for a black hole, the first law needs to be modified such that all dimension-full variables appearing in the theory also appear in the first law. Since the parameter \(l\) has dimensions of length, it should appear in the first law with its conjugate variable. Thus, the first law in this case takes the following form,
\begin{equation}
    dM=TdS+\psi dl~.
\end{equation}
Here, \(\psi\) is the variable conjugate to \(l\), which can be defined as 
\begin{equation}
\psi=\frac{\partial M}{\partial l}\bigg|_S.
\end{equation}
Under such modification, the physical meaning of \(M\) may not be associated to the internal energy \cite{kostor} but to some other thermal potential.\

\noindent The Smarr relation can be derived by a simple dimensional analysis for the black hole. For the Hayward black hole, \([M]=[L]\), \([S]=[L]^2\) and \([l]=[L]\). Here, \([L]\) denotes dimensions in length. Now, the Smarr relation for the black hole can be calculated using Euler's theorem for quasi-homogeneous functions, and it takes the following form,
\begin{equation}
    M=2TS+\psi l~.
\end{equation}
Thus, one may extend the thermodynamic phase space and study various properties related to it under such an extension. However, without loss of generality, we shall simply be setting the scale by choosing a value of the parameter \(l\) and study the thermal properties for such configurations. We are more interested on the impact of the parameter \(l\) on black hole thermodynamics. Next, we focus on the phase structure of the Hayward black hole.\\ 

\noindent \textbf{Heat Capacity and Free Energy}\\

\begin{figure}
    \includegraphics[width=1\linewidth]{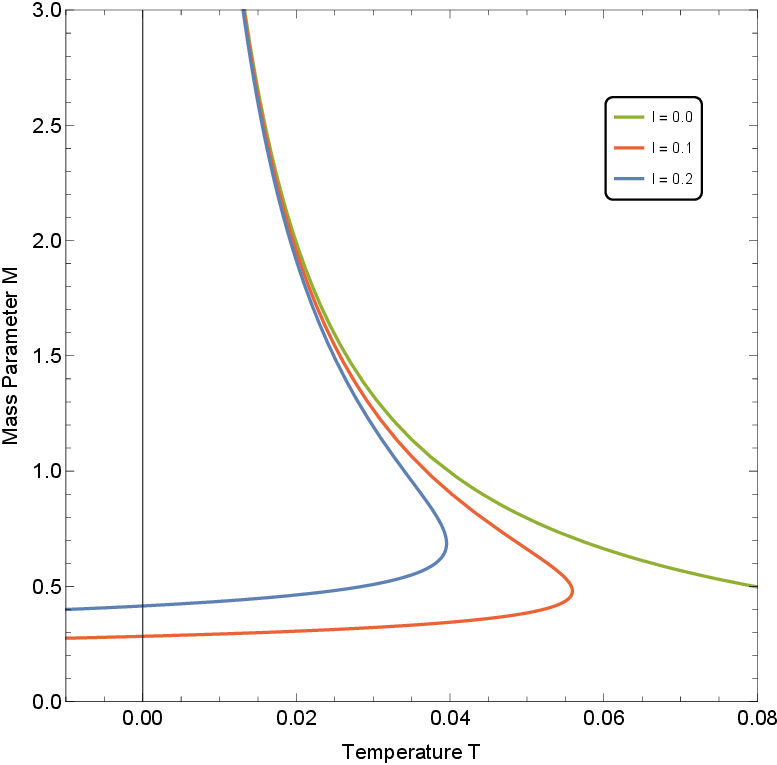}
        \caption{Parameter M vs Hawking Temperature T for \(l=(0, 0.1, 0.2)\). }
\label{masst}
    \end{figure}

\noindent Black holes show interesting phase structure and phase transitions of different kinds \cite{davis, HawkingPage, Chamblin1, Chamblin2, mann, neeraj1, neeraj2, neeraj3}. Usually, thermal stabilty and phase transition can be understood by calculating various response functions. Heat capacity is one such quantity which provides the information about the stability of different phases along with phase transition among different phases of black hole states. For the black hole under consideration, one can calculate the heat capacity using the formula 
\begin{equation}
    C(r_+)=\frac{dM}{dT}=\frac{dM/dr_+}{dT/dr_+}~.
\end{equation}
Although, it should be noted that the parameter \(M\) is not a single valued function of the Hawking temperature \(T\) for non-zero values of the Hayward parameter, \(l\). This is be depicted in Fig.(\ref{masst}). However, one can study the behaviour either for different branches separately, or with respect of the horizon radius as it is a better tuning parameter than the temperature itself.\
    
\begin{figure}
    \includegraphics[width=1\linewidth]{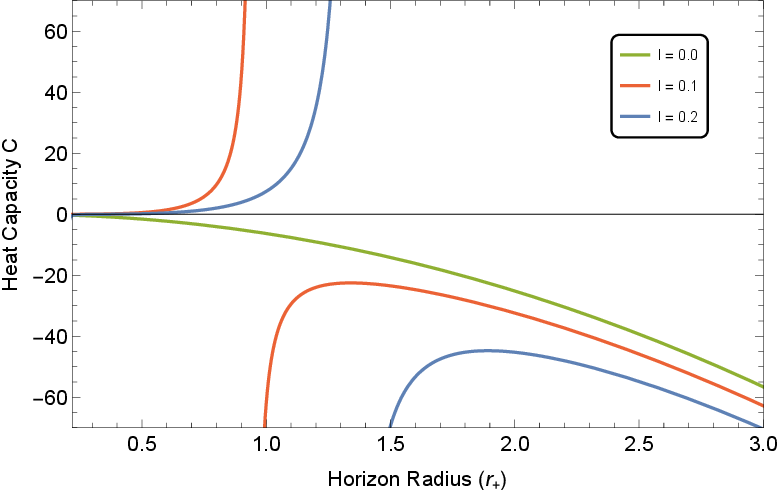}
        \caption{Heat Capacity vs Horizon Radius for \(l=(0, 0.1, 0.2)\). }
\label{heatcap}
    \end{figure}

\noindent Using eq.(s)(\ref{temp}, \ref{mass}), the expression for the heat capacity in terms of the horizon radius becomes,
\begin{equation}
    C=\frac{2\pi r_+^6(r_+^2-3l^2)}{(r_+^2-l^2)^2(9l^2-r_+^2)}~.
\end{equation}
It is clear from the above expression that the heat capacity diverges at two values of the horizon radii, \(r_+=l\) and \(r_+=3l\). However, the singular point, \(r_+=l\) corresponds to a black hole below the extremal limit (\(r_+=\sqrt{3}l\)). Thus, there is only one physically allowed singular point where the heat capacity diverges for non-zero values of \(l\). This point corresponds to the maxima in the temperature plot, given in fig.(\ref{f1}). This behaviour is similar to the Kerr-Newman black hole \cite{davis}. We plot the heat capacity with horizon radius for different values of parameter \(l\) in Fig.(\ref{heatcap}).
For the limit \(l\rightarrow 0\) (that is, for the Schwarzschild black hole), the heat capacity is negative throughout the range of \(r_+\) as shown in the figure. It is also apparent from the same figure that the small black holes are thermally stable (that is, they have a positive heat capacity) while the large black holes thermally unstable for non-zero values of \(l\). This behaviour can also be understood from the Hawking temperature plot in fig.(\ref{f1}). \

\begin{figure}
    \includegraphics[width=1\linewidth]{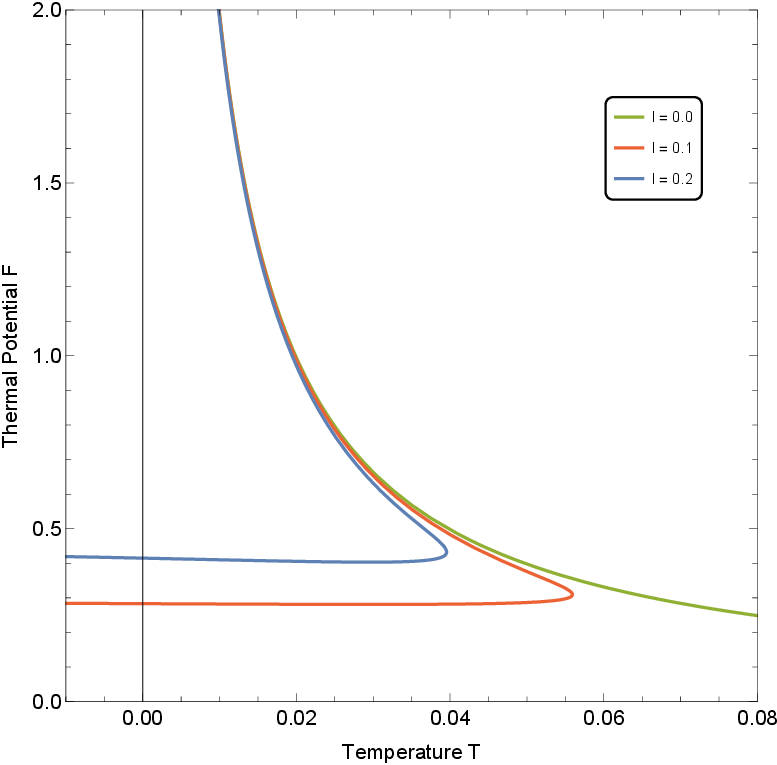}
        \caption{Thermal Potential F vs Temperature T for \(l=(0, 0.1, 0.2)\). }
\label{free}
    \end{figure}

\noindent Next, we carry out free energy analysis to study the phase transition associated with the singularity in the heat capacity. We consider an ensemble which is characterized by a potential defined as, 
\begin{equation}
    F=M-TS~.
\end{equation}
This is similar to canonical ensemble where \(F\) represent Helmholtz free energy. For the black hole under consideration the expression of the potential takes the following form,
\begin{align}
    F= & \frac{r_+^3}{2(r_+^2-l^2)}-\frac{1}{4}\left[\frac{1}{r_+}-\frac{3l^2}{r_+^3}\right]\nonumber\\
    &\times\left(r_+^2+2l^2\log{\frac{r_+^2-l^2}{l^2}}-\frac{l^2(r_+^2-l^2 +1)}{r_+^2-l^2}\right)~.
    \label{freepot}
\end{align}
This potential is plotted in Fig.(\ref{free}) with Hawking temperature for different values of the Hayward parameter. 
It shows that for non-zero values of \(l\), the free energy is also a multi-valued function of temperature with the maximum value of the possible temperature connecting two branches. Another important observation regarding the plot is its smooth nature. It implies that there is no first order phase transition in this system. Thus, one has to look for higher order phase transition depicted in the heat capacity plot. However, starting with this multi-valued potential expression in eq.(\ref{freepot}), one cannot proceed further to analyse higher order phase transitions from standard single-valued free energy perspectives. This would require some non-trivial analysis near the singular point, which we shall address in some future work. We shall rather explore implications of the novel entropy formula, eq.(\ref{new_ent}), to constrain the  final black hole parameters under the equal mass binary black hole merger scenario.    

\section{Constraints on Black Hole Merger}

The entropy formula, obtained in this work, while respecting the first law for the black hole is very different from the Bekenstein's area law. Along with the well known log correction term, usually appearing near the extremal black hole limit, to the standard area term, there appears one more term in this case, which might have some non-trivial implications. In this section, we shall consider merger of two equal mass Hayward black holes in asymptotically flat background and study the possible post-merger bounds on the parameter \(M\), which is linked to the mass of the resulting black hole. Similar to the  Hawking's analysis \cite{Hawkingmerger} for the case of standard black holes, we also consider the validity of the second law of thermodynamics. This provides an inequality, which renders bounds on the final black hole parameters. \

\begin{figure}[h!]
    \includegraphics[width=1\linewidth]{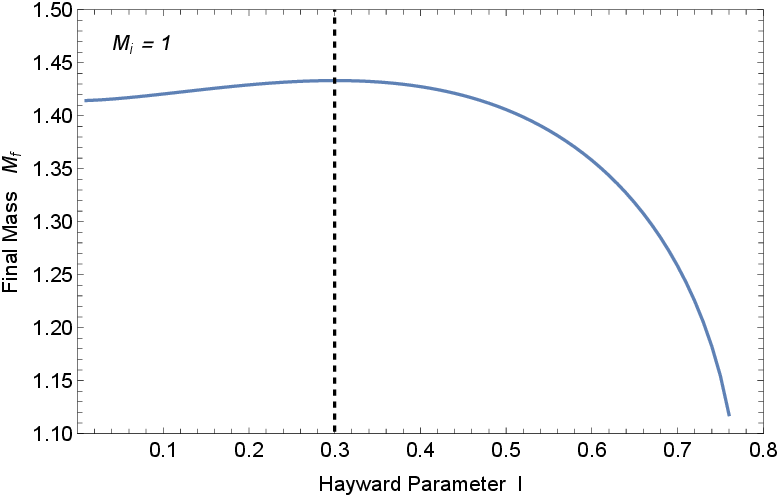}
        \caption{Final Mass Parameter \(M\) vs Hayward Parameter \(l\) for \(M_i=1\). }
\label{massl}
    \end{figure}

\noindent Now we shall start with considering two equal mass Hayward black holes such that the total entropy of this thermal system is \(2S_i\). Here \(S_i\) is the entropy of the single black hole with mass parameter \(M\). The expression of \(S_i\) takes the following form,
\begin{equation}
    S_i=\pi\left(r_+^{i2}+2l^2\log{\frac{r_+^{i2}-l^2}{l^2}}-\frac{l^2(r_+^{i2}-l^2 +1)}{r_+^{i2}-l^2}\right)~.
\end{equation}
Here, \(r_+^i\) is the horizon radius of the pre-merger black hole. We consider the head-on collision scenario of these black holes such that the post-merger black hole has the horizon radius \(r_+^f\) and the entropy \(S_f\). If this merger event has to follow the second law of black hole thermodynamics then we have the following inequality must hold,
\begin{equation}
    S_f\geq 2S_i~.
\end{equation}
which may be explicitly written in terms of $r_+^i , r_+^f$ and $l$ as, 
\begin{align}
   & \left(r_+^{f2}+2l^2\log{\frac{r_+^{f2}-l^2}{l^2}}-\frac{l^2(r_+^{f2}-l^2 +1)}{r_+^{f2}-l^2}\right)\nonumber\\
   &\geq 2\left(r_+^{i2}+2l^2\log{\frac{r_+^{i2}-l^2}{l^2}}-\frac{l^2(r_+^{i2}-l^2 +1)}{r_+^{i2}-l^2}\right)~.
\end{align}
This provides a bound on the maximum possible horizon radius for the black hole post-merger for given initial black holes. This in turn puts a bound on the mass parameter \(M\). We plot the final black hole mass parameter \(M\) with the Hayward parameter \(l\) in fig.(\ref{massl}) for the initial mass parameter \(M_i=1\), in appropriate units. 
The plot has several interesting features. The point where the curve meets the \(M_f\)-axis represents a minimum bound on the final mass parameter for the Schwarzschild black hole case. As one increases the Hayward parameter values, the bound becomes more stringent (that is, a less wider allowed mass range) and then decreases after reaching a certain maximum value. This is shown with the black dotted line in fig.(\ref{massl}). Notice that such a behaviour has also been observed in a recent study \cite{neeraj4} for another regular black hole merger case.\

\begin{figure}[h!]
    \includegraphics[width=1\linewidth]{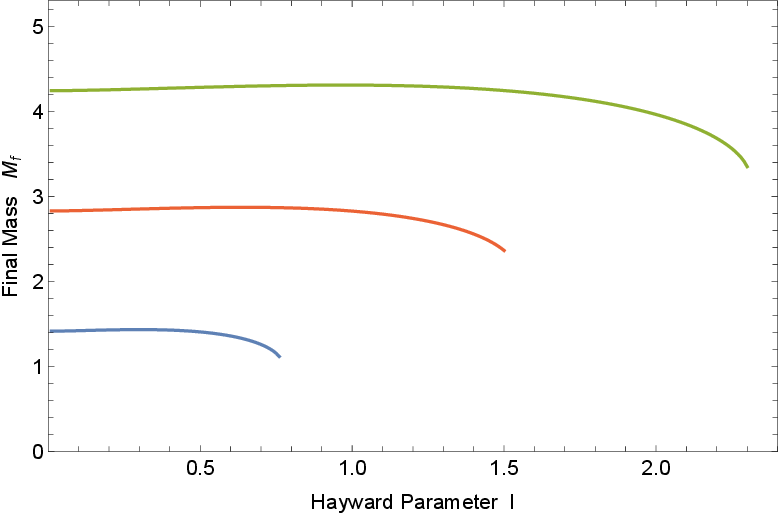}
        \caption{Final Mass Parameter \(M\) vs Hayward Parameter \(l\) for \(M_i=\textcolor{mmaBlue}{1}, \textcolor{mmaOrange}{2}, \textcolor{mmaGreen}{3}\). }
\label{massl1}
    \end{figure}

\noindent Next, we studied the impact of the Hayward parameter on the final mass parameter for different initial black hole mass parameters. The qualitative behaviour remains same as the initial mass parameter values, \(M_i\), are increased. The restrictions on the \(l\) values in fig.(\ref{massl1}) are due to extremal black hole limit for the initial black hole mass parameter \(M_i\). The important feature to note is that towards the maximum allowed values of parameter \(l\), the bound on the final mass parameter \(M_f\) reduces significantly. This observation is more apparent in fig.(\ref{massl}).  

\section{Discussions}

\noindent Among regular black hole solutions, the Hayward black holes hold special importance from the dynamical perspective. Even with one asymptotic charge these black holes have very interesting classical and thermal characteristics. Classically, their causal structure is similar to the charged black holes as these have an inner and an outer horizon. Along with that there exits an extremal limit as well. Keeping these interesting properties in mind, we explored its thermal and merger properties in a new thermal phase space set-up. We assumed that the thermodynamic laws hold for these objects and an aysmptotic observer observes Hawking temperature as a temperature associated with the event horizon. We also considered that the parameter \(M\) still represents the internal energy of the black hole system and the first law holds in this case. Under these reasonable assumptions, we derived a new entropy formula which has an interesting form. Along with the standard area term, there appears a well known log correction term and another term inversely proportional to the area. Interestingly, these log correction terms, which otherwise appear only in low temperature limit (that is, near extremal limit), appear at all temperatures. Another interesting property of these black holes is observed from the Hawking temperature plots. This shows that there is a maximum temperature value attained by the black hole as the size is varied. The plot provides information related to black hole stability as well. For smaller black holes, the temperature increases as the size increases, thus, rendering them stability in a canonical ensemble. Large black holes, on the other hand, are thermally unstable.    \ 

\noindent Next, we studied properties related to the phase structure and thermal stability of these Hayward black holes through response functions and free energy. We calculated the heat capacity and plotted it against the horizon radius as it is a single-valued function of this variable. We found that there is a point of discontinuity where the heat capacity diverged. This is similar to Davies point in Kerr-Newman black holes. Complementing to the observations in the temperature plot, the heat capacity is positive for the small black holes and negative for the large ones. \ 

\noindent We have also considered studying the phase transition of these black holes. We obtained an expression of a thermal potential similar to the free energy expression and plotted it with the Hawking temperature. We observed that the thermal potential is a multi-valued smooth function of temperature. This indicates that the phase transition is of order higher than one. We left this non-trivial analysis near the phase transition point for future work. \ 

\noindent Apart from the thermal impact of the novel entropy formula, it also constrains all gravitational events when the second law of black hole mechanics/thermodynamics are considered valid. We considered how this law, under this new entropy formula, constrains the binary merger event. We considered head-on collision of two equal mass Hayward black holes. We plotted the final black hole mass parameter with the Hayward parameter for a fixed value of the initial mass parameter. We observed that there is a specific value of the Hayward parameter for which the bound on the final mass parameter is the strongest (that is, a narrower allowed mass range). We also studied the impact of change in the initial mass parameter on the final mass bound. We found that there is no qualitative change for different initial values. In terms of the maximum energy radiated to asymptotic infinity, the impact is most prominent near the maximum allowed \(l\) values which is constrained by the initial mass parameter value.\ 

\noindent The Hayward parameter is a phenomenologically introduced parameter, which might have its origin in quantum gravity. At empirical level, this might have noticeable impacts on the properties of the gravitational systems, in principle. This parameter also solves the singularity problem and addresses the information paradoxes in gravity. Thus, it should be important to see its impacts on the experimental and theoretical fronts to get potential hints about the underlying quantum theory gravity.

\section*{Acknowledgement} This research has received funding support from the NSRF via the Program Management Unit for Human Resource and Institutional Development, Research and Innovation grant number B13F680075. AS would like to acknowledge Mrs. Megha Dave for financial support during the work.

\setlength{\emergencystretch}{3em}
\balance

\end{document}